\documentclass[fullpage]{article}
\usepackage{fullpage}
\usepackage{graphicx,amssymb,amsfonts,amsmath,amsthm}
\usepackage{cite,hyperref,subfigure}

\newtheorem{prop}{Proposition}
\newtheorem{thm}{Theorem}

\newtheorem{lemma}{Lemma}

\def \1{\mathbf 1}
\def \0{\mathbf 0}

\def \E{\mathbb E}

\def \B{\mathbb B}
\def \x{\mathbf x}
\def \z{\mathbf z}
\def \y{\mathbf y}
\def \b{\mathbf b}
\def \Tc{\mathcal T}

\def \L{\mathcal L}
\def \N{\mathcal N}
\def \H{\mathcal H}
\def \C{\mathcal C}

\date{}

\begin{document}

\title{Detecting Weak but Hierarchically-Structured Patterns in Networks}
\maketitle

\author{
\begin{center}
\begin{tabular}{c}
Aarti Singh\\
Machine Learning Department\\
Carnegie Mellon University\\
\texttt{aartisingh@cmu.edu}\vspace{0.2cm}\\
Robert D.~Nowak\\
Department of Electrical and Computer Engineering\\
University of Wisconsin - Madison \\
\texttt{nowak@engr.wisc.edu}\vspace{0.2cm}\\
Robert Calderbank\\
Department of Electrical Engineering\\
Princeton University \\
\texttt{calderbk@princeton.edu}\\
\end{tabular}
\end{center}
}

\begin{abstract}
The ability to detect weak distributed activation patterns in networks is critical to several applications, such as identifying the onset of anomalous activity or incipient congestion in the Internet, or faint traces of a biochemical spread by a sensor network.  This is a challenging problem since weak distributed patterns can be invisible in per node statistics as well as a global network-wide aggregate. Most prior work considers situations in which the activation/non-activation of each node is statistically independent, but this is unrealistic in many problems.  In this paper, we consider structured patterns arising from statistical dependencies in the activation process.  Our contributions are three-fold. First, we propose a sparsifying transform that succinctly represents structured activation patterns that conform to a hierarchical dependency graph.  Second, we establish that the proposed transform facilitates detection of very weak activation patterns that cannot be detected with existing methods. Third, we show that the structure of the hierarchical dependency graph governing the activation process, and hence the network transform, can be learnt from very few (logarithmic in network size) independent snapshots of network activity.
\end{abstract}

\section{Introduction}
We consider the problem of detecting a weak binary pattern corrupted by noise
that is observed at the $p$ nodes of a network:
$$
y_i = \mu x_i + \epsilon_i \hspace{0.2cm} i = 1,\dots, p
$$
Here $y_i$ denotes the observation at node $i$ and $\x = [x_1,\dots,x_p] \in \{0,1\}^p$ is the $p$-dimensional {\em unknown} binary activation pattern, $\mu >0$ denotes an {\em unknown} signal strength, and the noises $\{\epsilon_i\} \overset{\rm iid}{\sim} {\cal N}(0,\sigma^2)$, the Gaussian distribution with mean zero and variance $\sigma^2$. The condition $x_i = 0$, $i=1,\dots,p$, is the baseline or normal operating condition (no signal present).  If $x_i>0$ for one or more $i$, then a signal or activation is present in the network. We are interested not in arbitrary patterns of activation, but rather our focus is on patterns that are related to the physical structure of the network and/or to other statistical dependencies in the signal. This is motivated by problems arising in practice, as discussed below.  More specifically, we consider classes of patterns that are supported over hierarchically-structured groups or clusters of nodes. Such a hierarchical structure could arise due to the physical topology of the network and/or due to dependencies between the nodes. For example, hierarchical dependencies are known to exist in gene networks due to shared regulatory pathways \cite{gene_tree,socbio_tree}, empirical studies show that Internet path properties such as delay and bandwidth are well-approximated by tree-embeddings \cite{internet_tree}, sensor networks are often hierarchically structured for efficient management \cite{sens_tree}, and communities in social networks can be hierarchical \cite{socbio_tree}.  We address the problem of detecting the presence of weak but structured activation patterns in the network. This problem is of interest in several applications including detecting incipient congestion or faint traces of malicious activity in the Internet, early detection of a chemical spread or bio-hazard by a sensor network, identification of differentially expressed set of genes in microarray data analysis, or malicious groups in social networks.

If $\x$ is known, then the optimal detector is based on
aggregating the measurements of the locations known to contain the signal (e.g., in the classical distributed detection literature it is often assume that $x_i = 1$ for all $i$ or
$x_i=0$ for all $i$ \cite{varshney}).  We are interested in cases where $\x$ is unknown.
If $\x$ is arbitrary, this is a problem in literature known as the {\em multi-channel
signal detection problem} \cite{Ingster}. In this case, global aggregation rule (testing the average
of all node measurements) can reliably detect any signal strength $\mu > 0$ if the number of active
locations $\|\x\|_0 > \sqrt{p}$. This is because $\frac1{\sqrt{p}}\sum^p_{i=1}y_i \sim \N\left(\mu \frac{\|\x\|_0}{\sqrt{p}},\sigma^2\right)$, and therefore as the network size $p$ grows, the
probability of false alarm and miss can be driven to zero by choosing an appropriate threshold.
However, in the high-dimensional setting when $p$ is very large and the activation is
sparse $\|\x\|_0 \leq \sqrt{p}$, then different approaches to detection are required. If the signal strength
$\mu > \sqrt{2\sigma^2\log p}$, then the signal can be reliably detected using the max statistic
$\max_i y_i$, irrespective of the signal sparsity level. This is because if there is no signal,
the max statistic due to noise alone (maximum of $p$ iid $\N(0,\sigma^2)$ random variables)
is $\leq \sqrt{2\sigma^2\log p}$ with probability 1, in the large $p$ limit. Therefore,
the most challenging case is when the network activation is
\begin{table}[h]
\centering
\begin{tabular}{l l}
{\em weak:}  $\mu < \sqrt{2\sigma^2\log p}$ \ \ \ and \ \ \ {\em sparse:}  $\|\x\|_0 < \sqrt{p}$
\end{tabular}
\end{table} \\ \vspace{-.3in} \\
\noindent In this case, the signal is buried in noise and cannot be detected in per node measurement
or in global network-wide aggregate. This necessitates {\em selective and adaptive fusion}
where the node measurements to be aggregated are chosen in a data-driven fashion.
One approach that is common in the signal processing literature is to consider the generalized
likelihood ratio test (GLRT) statistic $\max_{\x\in\{0,1\}^p} \x^T\y/\x^T\x$
where the observed vector is matched with all $2^p$ possible true activation patterns.
However, in high-dimensional settings, the GLRT is computationally intractable.
For weak and sparse signals, the limits of detectability were
studied by Ingster \cite{Ingster}, and subtle tests that are adaptive in various ranges of
the unknown sparsity level were investigated. More recently, test statistics have been proposed
 \cite{Jin-Donoho,Jager-Wellener} that can attain the detection boundary
simultaneously for any unknown sparsity level. A generalization of this problem has also been
studied in \cite{Tsy_spclassb}.
However, all the work above assumes that the activations at nodes are independent of each other.
As a result, the signal strength $\mu$ must be $> c \sqrt{\log p}$ for some constant $c >0$ and
hence the signal cannot be too weak.

The assumption of independent activations is often unreasonable in a network setting, where the observations at nodes tend to be highly dependent due to the structure of the network and/or dependencies in the activation process itself. For example, routers in the same autonomous system will show similar variations in round-trip-time measurements, or co-located sensors monitoring an environmental phenomena will have correlated measurements.  Recently, there has been some work aimed at structured patterns of activation in graphs \cite{Castro-Donoho,Castro-Emmanuel,Castro-Emmanuel-Arnaud}, which indicates that it is possible to detect even weaker signals
by leveraging the statistical dependencies in the activation process.  
Of these, the lattice-based models in \cite{Castro-Emmanuel-Arnaud} are most closely related to our work, but they do not capture the hierarchical structure we have in mind, nor do they appear to offer a computationally tractable approach to detection.  We also mention the recent work of \cite{Addario-Berry}, which establishes
fundamental limits of detectability for several classes of structured patterns in graphs. The detection tests proposed in that paper are generally combinatorial in nature (like the GLRT mentioned above), requiring a brute-force examination of all patterns in each class, and therefore are computationally prohibitive in all but very low-dimensional situations.

In this paper, we consider a different class of patterns that reflects the hierarchical dependencies present in many real-world networks and leads to computationally practical detection methods.  Furthermore, we demonstrate that it is possible to learn the hierarchical dependency structure of the class from a relatively small number of observations, adding to the practical potential of our framework.  The hierarchical dependencies structures we consider tend to results in network activation patterns that are supported over hierarchically-organized groups or clusters of nodes.  We will show that such {\em structured} activation patterns can be sparsified even further by an orthonormal transformation that is adapted to the dependency structure. The transform concentrates the unknown $\x$ in a few large basis coefficients, thus facilitating detection. We show that if the canonical domain sparsity $\|\x\|_0 \sim p^{1-\alpha}$ and the transform domain sparsity scales as $p^{1-\beta}$, where $\beta > \alpha$, then the threshold of detection scales as $\mu > p^{-(\beta-\alpha)/2}\sqrt{2\sigma^2\log p}$. Contrasting this with the detectability threshold of earlier methods $\mu >\sqrt{2\eta_\alpha\sigma^2\log p}$ \cite{Ingster,Jin-Donoho} (where $0<\eta_\alpha<1$ is independent of $p$), we see that a polynomial improvement is attained if the activation pattern is sparser in the transform domain.
Hence, by exploiting the hierarchial structure of $\x$, we can detect extremely faint activations that could not be detected using existing methods.

Our contributions are three-fold.  First, we propose a sparsifying transform based on hierarchical clustering that is adapted to the dependency structure of network measurements. We propose a practically-motivated  generative model that allows for arbitrary activation patterns, but favors patterns that are supported over hierarchically-organized groups of nodes. We show that patterns from this model are  compressed by the sparsifying transform. Though we focus on the detection problem in this paper, the sparsifying transform could be exploited in other problem domains, e.g. de-noising, compression, sparse regression, variable selection, etc.
Second, we establish that the sparsifying transform can amplify very weak activation patterns by effectively performing adaptive fusion of the network measurements. Since the network activity is summarized in a few large transform coefficients, the signal-to-noise ratio (SNR) is increased, and this facilitates detection of very weak activation patterns. We quantify the improvement in the detection threshold relative to existing methods. The detection method we propose is a constructive procedure and computationally efficient. Third, we do not necessarily assume that the graph structure is known a priori, and show that the dependency structure, and hence the sparsifying transform, can be learnt from very few, $O(\log p)$, multiple independent snapshots of network measurements.

The rest of this paper is organized as follows. In section~\ref{sec:hier_trans}, we introduce the sparsifying transform. We propose a generative model in Section~\ref{sec:spar_det} for hierarchically-structured patterns, and characterize the sparsifying properties
and detection threshold attained by the proposed transformation.
 Section~\ref{sec:learn_hier} examines the sample complexity of learning the hierarchical dependencies
 and transform from data. Simulations 
 are presented in Section~\ref{sec:sim}. 
 Proofs sketches are given in the Appendix.

\section{Hierarchical structure in Networks}
\label{sec:hier_trans}
As discussed in the introduction, activation patterns in large-scale networks such as the Internet, sensor, biological and social networks often have hierarchical dependencies. This hierarchical dependency structure can be exploited to enable detection of very weak and sparse patterns of activity. In this section, we propose a transform that is adapted to a given set of pairwise similarities between nodes.  The similarity of node $i$ and $j$ is denoted by $r_{ij}$. For example, $r_{ij}$ could be the covariance between measurements at node $i$ and $j$, but other similarity measures can also be employed.  The transform is derived from a hierarchical clustering based on the similarity matrix $\{r_{ij}\}$.  If the matrix reflects an underlying hierarchical dependency structure, then the resulting transform sparsifies activation patterns supported on hierarchically-organized groups of nodes.

\subsection{Hierarchical Clustering of Nodes} We employ a standard, bottom-up agglomerative clustering algorithm. The algorithm takes as input a set of pairwise similarities $\{r_{ij}\}$ and returns a hierarchical set of clusters/groups of nodes, denoted as $\H$.  The algorithm is described in Figure ~\ref{tab:clustering_algo}. Suppose instead that we are given a hierarchical set of clusters $\H^*$.  What conditions must a similarity matrix satisfy, in relation to $\H^*$, so that the agglomerative clustering algorithm recovers $\H^*$ and not some other hierarchical clusters?  This is an important question for several reasons as we will see in subsequent sections (e.g., to robustly identify $\H^*$ from a noisy observation of the similarity matrix).  To answer this question first note that the agglomerative clustering algorithm always merges two clusters at each step. Therefore, the most we can hope to say is that under some conditions on the similarity matrix, the agglomerative clustering algorithm produces a hierarchical set of clusters $\H$, such that $\H^* \subset \H$; i.e., $\H$ contains all cluster sets in $\H^*$, but may include additional subsets due to the restriction of binary merging.
 The following lemma gives a sufficient condition on the similarity matrix to guarantee that this is the case. The proof is straightforward and omitted to save space.
\begin{lemma}
\label{lem:clus}
Suppose we are given a collection of hierarchical clusters $\H^*$.
If for every pair of clusters $(c,c') \in \H^*$, where $c' \subset c$, the
maximum similarity between any $i \in c'$ and $j \in c/c'$ is smaller than
the minimum similarity between any pair of nodes in $c'$, then
the agglomerative clustering algorithm of Figure~\ref{tab:clustering_algo} recovers $\H^*$.
\end{lemma}

\subsection{Hierarchical Basis for Network Patterns}
Based on a hierarchical clustering of network nodes, we propose the following unbalanced Haar basis representation for activation patterns. When two clusters $c_1$ and $c_2$ are merged in the agglomerative clustering algorithm, a normalized basis vector is defined (up to normalization) by
$$
 \mathbf{b} \propto \frac1{|c_2|}\mathbf{1}_{c_2}- \frac1{|c_1|}\mathbf{1}_{c_1},
$$
 where $\mathbf{1}_{c_i}$ denotes the indicator of the support of subcluster $c_i$.  Projecting the activation pattern $\x$ onto this basis vector computes a difference of the average measurement on each constituent cluster. As a result, the basis coefficient $\mathbf{b}^T\x$ is zero if the nodes in the constituent clusters are all active or inactive. Thus, the basis vectors possess one vanishing moment akin to standard Haar wavelet transform, and will sparsify activation patterns that are constant over the merged clusters.  This procedure yields $p-1$ difference basis vectors. These basis vectors are augmented with the constant vector that computes the global average.  The resulting vectors form the columns of an orthonormal unbalanced Haar transform matrix $\B$.

The proposed method of hierarchical clustering followed by basis construction is similar in spirit to the recent work of Lee et al. \cite{Treelets} on treelets and of Murtagh \cite{Murtagh}.  However, treelets do not lead to a sparsifying transform in general if the node measurements or aggregates have different variances. The work of Murtagh uses balanced Haar wavelets on a dendrogram and does not yield an orthonormal basis since the basis vectors are not constant on sub-groups of nodes. As a result, the transform coefficients are correlated and dependent, making the resulting statistics difficult to analyze. Our procedure, on the other hand, is based on {\em unbalanced} Haar wavelets which are constant on sub-groups of nodes and thus result in orthogonal vectors.

\begin{figure}[ht] \begin{center} \framebox{ \begin{minipage}{.7\textwidth}
{\em Input:} Set of all nodes $\L=\{1,\dots,p\}$ and pairwise similarities $\{r_{ij}\}_{i,j\in\L}$\\ \vspace{-.1in} \\
{\em Initialize:} Clusters $\C = \{\{1\},\{2\},\dots,\{p\}\}$, \\Hierarchical clustering $\H = \C$, Basis $\B = [\ ]$\\ \vspace{-.1in} \\
{\bf while} $|\C|>1$ \vspace{-.1in}
\begin{itemize}
\item[] Select $(c_1,c_2) = \arg\max_{c_1,c_2 \in \C}  \frac{\sum_{i \in c_1} \sum_{j \in c_2} r_{ij} }{|c_1| |c_2|}$ 
\item[] Merge  $c = c_1\cup c_2$
\item[] Update
\begin{tabular}{l l }
$\H = \H \cup \{c\}$\\
$\C = (\C / \{c_1,c_2\}) \cup \{c\}$
\end{tabular}
\item[] {\em Construct unbalanced Haar basis vector:}\\\vspace{-0.2in}
        \begin{eqnarray*}\b & = &\frac{\sqrt{|c_1||c_2|}}{\sqrt{|c_1|+|c_2|}}\left[\frac1{|c_2|}\1_{c_2}-
        \frac1{|c_1|}\1_{c_1}\right]\\
        \B & = & [\B |\ \b]
        \end{eqnarray*}
\end{itemize} \vspace{-.2in}
{\bf end}\\ \vspace{-.1in} \\
$\b = \frac1{\sqrt{|\L|}}\1_{\L}$, $\B = [\B |\ \b]$\\\\
{\em Output:} $\B$, $\H$
\end{minipage}
}
\end{center}
\caption{Algorithm for hierarchical clustering.}
\label{tab:clustering_algo}
\end{figure}

\subsection{Activations of Hierarchically-Organized Groups}
To illustrate the effectiveness of the proposed transform, consider activation patterns
generated by the union of a small number of clusters of the hierarchical collection $\H$,
i.e. let $\x = \1_{\cup^m_{i=1}c_i}$, where $c_i\in \H$.
Then it is not difficult to see that the transform of $\x$ will produce no more than $O(m)$ non-zero basis coefficients. The magnitude of each coefficient will be proportional to the square-root of the number of nodes in the corresponding cluster on which the basis is supported. Suppose that the largest cluster contains $k$ nodes. Then the largest coefficient of $\x$ will be on the order of $\sqrt{k}$. This implies that the corresponding coefficient of the noisy observations $\y$ will have a signal-to-noise energy ratio (SNR) of order $k/\sigma^2$, compared to the per node SNR of $1/\sigma^2$ in the canonical domain, making the activation much more easily detectable.

In practice, actual activation patterns may only approximate this sort of ideal condition, but the transform can still significantly boost the SNR even when the underlying activation is only approximately sparse in the transform domain. In the next section we propose a practically-motivated generative model capable of generating arbitrary patterns. As the parameter of the model is varied, the patterns generated from the model tend to have varying degrees of sparseness in the transform domain.

\section{Sparsifying and Detecting Activations}
\label{sec:spar_det}
In this section, we study the sparsifying capabilities of the proposed transform,
and the corresponding improvements that can be attained in the detection threshold.
For this, we introduce a generative model that, with high probability,
produces patterns that are approximately sparse.

\subsection{A Generative Model for Activations}
We model the hierarchical dependencies governing the activation process by a multi-scale
latent Ising model, defined as follows. Let $\Tc^* = (V,E)$ denote a tree-structured graph
with $V$ as the vertex set and $E$ as the edge set. For simplicity,
we assume that the degree of each node is uniform, denoted as $d$,
and let $L = \log_d p$ denote the depth of the tree. The leaves $\L$ of the tree are
at the deepest level $L$ and correspond to the network nodes, while the internal vertices characterize the multi-scale
dependencies between the node measurements. Let $\z$ denote a $|V|$-dimensional
vector of variables defined over the complete tree, but we only observe $\x = \{z_i\}_{i\in\L}$, the $p$-dimensional
vector of network observations. We assume that $\z$ (and hence $\x$) is generated according
to the following probabilistic Ising model:
\begin{equation*}
p(\z) \propto \exp\left(\sum^L_{\ell=1} \gamma_\ell \sum_{i\in V_\ell}[z_i z_{\pi(i)} + (1-z_i)(1-z_{\pi(i)})]\right)
\end{equation*}
Here $V_\ell$ denotes the vertices at level $\ell$,
and $\gamma_\ell>0$ characterizes the strength of pairwise interaction between a vertex $i$ at level
$\ell$ and its parent $\pi(i)$. This model implies that the $2^p$ possible
activation patterns are not equiprobable, and the probability of a pattern is higher if the
variables agree with their parents in the tree dependency graph $\Tc^*$. This is a natural model for several application
domains where the activation is governed by a contact process, e.g. the spread of an infection or disease.

\subsection{Canonical and Transform Domain Sparsity}
To evaluate the transform domain sparsity, we first establish that the latent tree
dependency graph $\Tc^*$ can be recovered by the agglomerative hierarchical clustering algorithm
of Figure~\ref{tab:clustering_algo}. Based on a result by Falk \cite{Falk75},
the covariance between any two leaf variables $i$ and $j$ is proportional to
$\Pi^L_{\ell = \ell'+1} (\tanh \ \gamma_\ell)^2$, where $\ell'$ denotes the level of the
root of the smallest subtree containing $i$ and $j$ (i.e. smallest cluster containing $i$ and $j$).
Thus, if the covariance is used as the similarity measure, it is easy to verify that it satisfies the conditions of Lemma~\ref{lem:clus}. This is important since the covariance
could be estimated from observations of the network.
We have the following result.
\begin{prop}
The agglomerative hierarchical clustering algorithm of Figure~\ref{tab:clustering_algo} perfectly recovers the
tree-structured dependency graph $\Tc^*$ on which the Ising model is defined, when using covariance between the
leaf variables as the similarity measure.
\end{prop}

We now show how the unbalanced Haar basis built on the tree dependency graph $\Tc^*$ leads
to a sparse representation of binary patterns drawn from the multi-scale Ising model. Recall that a
transform coefficient is zero if the activation pattern is constant over the support of
the corresponding basis vector.
\begin{thm} Consider a pattern $\x$ drawn at random from a latent Ising model on a tree-structured graph
with uniform degree $d$ and depth $L = \log_d p$, as described in the previous section. If the interaction strength
scales with the level $\ell$ as $\gamma_\ell = \ell\beta\log d$ where $0\leq \beta\leq 1$, then with probability $> 1-\delta$,
the number of non-zero transform coefficients are bounded by
$$
\|\B^T\x\|_0 \leq 3d(\log_d p)^2 p^{1-\beta}. 
$$
for $p$ large enough. 
\label{thm:trans_sparsity}
\end{thm}
Proof is given in the Appendix.
Since the interaction strength increases with level, variables at deeper levels are less
likely to disagree with their parents and hence activation patterns supported over groups of
nodes are favored. The above theorem states that, with high probability, patterns generated
by this model are approximately sparse in the proposed transform domain. The degree of sparsity
is governed by $\beta$, the rate at which the interaction strength increases with level.

We also have in mind situations in which the number of total activations in the network
is small, i.e., $\|\x\|_0 < \sqrt{p}$, which renders the naive global
fusion test statistic unreliable (see discussion in Introduction).
To make widespread activations less probable, we constrain the Ising model as follows. Set the root vertex to the value $0$.  Let $\ell_0 = \frac{\alpha}{\beta}L$, where $0<\alpha<\beta$. Let $\gamma_\ell = \ell\beta\log d$ for $\ell \geq \ell_0$, and $\gamma_\ell = \infty$ 
for $\ell < \ell_0$. This model forces variables at scales coarser than $\ell_0$ to be identically $0$.  Proof of the following theorem is given in the Appendix.
\begin{thm} Consider a pattern $\x$ drawn at random from a latent Ising model on a tree-structured graph
with uniform degree $d$ and depth $L = \log_d p$. Let $\ell_0 = \frac{\alpha}{\beta}L$, where
$0<\alpha<\beta$, and the interaction strength
scale with the level $\ell$ as $\gamma_\ell = \ell\beta\log d$ for $\ell \geq \ell_0$, and
$\gamma_\ell = \infty$ 
for $\ell < \ell_0$. If the pattern
corresponds to the root variable taking value zero,
then with probability $> 1-4\delta$ and for $p$ sufficiently large,
the number of non-zero transform coefficients are bounded by
$$
\|\B^T\x\|_0 \leq 3 d(\log_d p)^2 p^{1-\beta}, 
$$
and the canonical domain sparsity is bounded as
$$
c p^{1-\alpha} \leq \|\x\|_0 \leq C (\log_d p) p^{1-\alpha}, 
$$
where $C > c>0$ are constant.
\label{thm:trans_can_sparsity}
\end{thm}
The result of the theorem states that the transform domain sparsity scales as $p^{1-\beta}$
(and is therefore determined by the rate at which the interaction strength increases with level),
while the canonical domain sparsity scales as $p^{1-\alpha}$ (and is therefore determined
by the smallest interaction strength between a variable and its parent). Since $\beta > \alpha$,
the proposed transform enhances the sparsity of canonically
sparse patterns that have a multi-scale group structure. In the next section, we show that
this enhanced sparsity implies a higher Signal-to-Noise (SNR) ratio in the transform domain,
thus facilitating detection.


\subsection{Threshold of Detectability}
Recall that the observed data is given by the following additive noise model:
$$
y_i = \mu x_i + \epsilon_i \hspace{0.2cm} i = 1,\dots, p
$$
where $\mu$ denotes the unknown signal strength, $\x$ is the unknown activation pattern, and
$\epsilon_i \stackrel{iid}\sim \N(0,\sigma^2)$. The detection problem corresponds to the following hypothesis test:
$$
H_0: \mu = 0  \mbox{\hspace{0.5cm} vs. \hspace{0.5cm}} H_1: \mu > 0
$$

Projecting the network data onto the basis vectors $\b \in \B$ yield the empirical transform coefficients
$\b^T_i\y$. If the pattern $\x$ is sparser in the transform domain, then its energy is concentrated in a
few non-zero coefficients. Thus, the signal-to-noise ratio is boosted and detection is easier.
To investigate the threshold of detectability for weak but structured activation patterns,
we consider a simple test based on the maximum of the absolute values of the
empirical transform coefficients $\max_i |\b_i^T\y|$ as the test statistic. The
following theorem provides an upper bound on the detection threshold using the max
statistic in the transform domain for patterns drawn from the tree-structured Ising model.
\begin{thm}
Consider a pattern $\x$ drawn at random from a latent Ising model on a tree-structured graph
with uniform degree $d$ and depth $L = \log_d p$. Let $\ell_0 = \frac{\alpha}{\beta}L$ and the interaction strength
scales with the level $\ell$ as $\gamma_\ell = \ell\beta\log d$ for $\ell \geq \ell_0$, and
$\gamma_\ell = \infty$ for $\ell < \ell_0$.

With probability $> 1-2\delta$ over the draw of the activation pattern, the test statistic
$\max_i |\b_i^T\y|$ drives the probability of false alarm and miss (conditioned on the draw
of the pattern) to zero asymptotically as $p \rightarrow \infty$ if the signal strength
$$
\mu > c \ p^{-\kappa}\sqrt{2\sigma^2\log p},
$$
where $\kappa = (\beta-\alpha)/2 > 0$ and $c>0$ is a constant.
\label{thm:detect}
\end{thm}
Proof is given in the Appendix. We see that a polynomial improvement is attained
if the activation pattern is sparser in a network transform domain. This is a significant improvement
over canonical domain methods that do not exploit the structure of patterns and
are limited to detecting signals with strength $\mu > \sqrt{2\eta_\alpha\sigma^2\log p}$
(where $0<\eta_\alpha<1$ is independent of $p$) \cite{Ingster,Jin-Donoho,Tsy_spclassb}.

\section{Learning Clusters from Data}
\label{sec:learn_hier}
In practice, the pairwise similarities or covariances used for hierarchical clustering and constructing the proposed transform can only be estimated from data. Since the empirical covariance between network nodes can be learnt from multiple i.i.d. snapshots of network measurements, we now provide finite sample guarantees on the recovery of the multi-scale dependency structure from empirically estimated covariances. Analogous arguments can also be made for any similarity measure provided the empirical estimates satisfy a concentration inequality.
\begin{thm}
\label{thm:emp_cov}
Consider noisy network measurements as per the following additive noise model:
$$
y_i = x_i + \epsilon_i \hspace{0.5cm} i = 1,\dots,p
$$
where $\epsilon_i$ are independent $\N(0,\sigma^2)$. The $x_i$ are independent
of the noise variables $\epsilon_i$, and are uniformly bounded by $M$.
For simplicity, we assume that the variables $x_i$ are also zero-mean.
Dependencies between the $\{x_i\}^p_{i=1}$ possess a hierarchical
structure. Specifically,
assume the covariances $\{\E[(x_ix_j)]\}$ satisfy the
conditions of Lemma~\ref{lem:clus} for a hierarchical set of clusters $\H^*$.
Let $\tau$ denotes the smallest difference (gap) between the minimum pairwise
covariance of leaf variables within any cluster and the maximum covariance between
leaf variables in different clusters. Also, let $r_{ij} = \E[(y_i y_j)] = \E[(x_i x_j)] + \sigma^2\delta_{ij}$,
where $\delta_{ij}$ is the Kronecker delta, denote the true covariance of the observed variables.
Notice that the noise only affects the auto-covariances which are irrelevant for clustering, and
hence $r_{ij}$ essentially behaves as $\E[(x_i x_j)]$ for clustering purposes.

Suppose we observe $n$ i.i.d noisy realizations $\{y^{(k)}_1,\dots,y^{(k)}_p\}^n_{k=1}$
of the $p$ leaf variables, and $\{\widehat r_{ij} = \frac1{n} \sum^n_{k=1}y^{(k)}_iy^{(k)}_j\}$
denote the empirical covariances. Let $\delta >0$. If
$$
\frac{n}{\log n} \geq \frac1{c_2\tau^2}\log(c_1 p^2/\delta),
$$
then with probability $>1-\delta$, the agglomerative clustering algorithm of
Figure~\ref{tab:clustering_algo} applied to $\{\widehat r_{ij}\}$ recovers $\H^*$.
Here $c_1, c_2 >0$ are constants that depend on $M$ and $\sigma^2$.
\end{thm}
Recall that $p$ denotes the number of network nodes. The theorem implies that only $O(\log p)$ measurements are needed to learn the hierarchical clustering and hence the proposed transform.

\section{Simulations}
\label{sec:sim}
\begin{figure}
\begin{center}
\includegraphics[scale=0.4, angle = -90, clip = true, viewport=0in 0in 9in 12in]{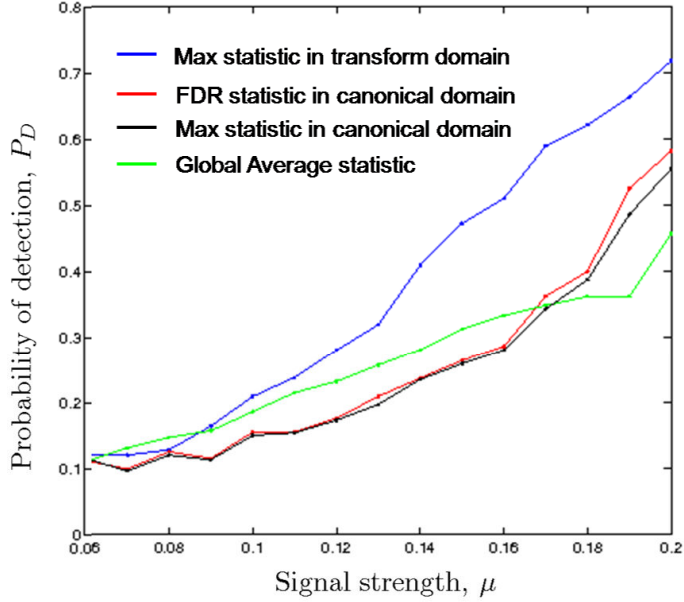}
\end{center}
\caption{Performance comparison of global fusion, FDR, and the max statistic in transform and canonical domains, for
weak patterns generated according to a hidden multi-scale Ising model.}
\label{fig:sparse_Ising}
\end{figure}

We simulated patterns from a multi-scale Ising model defined on a tree-structured graph with $p = 1296$ leaf
nodes with degree $d = 6$ and depth $L = 4$. The network observations are modeled by adding additive
white gaussian noise with standard deviation $\sigma = 0.1$ to these patterns. This implies that a weak
pattern is characterized by signal strength $\mu < \sigma\sqrt{2\log p} = 0.38$. We generate weak patterns
with signal strength $\mu$ varying from $0.06$ to $0.2$ and compare the detection
performance of the max statistic in transform and canonical domains,
and the global aggregate statistic, for a target false alarm probability of $0.05$.
We also compare to the FDR (False Discovery Rate) \cite{FDR_Benjamini} which is a canonical domain method that
orders the measurements and thresholds them at a level that is adapted to the unknown sparsity level.
The probability of detection as a function of signal strength is plotted in Figure~\ref{fig:sparse_Ising}.
Detection in the transform domain clearly
outperforms other methods since our construction exploits the network node interactions.

The algorithmic complexity of hierarchical clustering $p$ objects is $O(p^2 \log p)$, which essentially dominates the
complexity of the detection procedure we propose.


\section*{Appendix}


\subsection*{Proof of Theorem~\ref{thm:trans_sparsity}}
Each unbalanced Haar basis vector $\b \in\B$ (except for the global summary vector $\1_{\L}/\sqrt{|\L|}$)
has one vanishing moment, i.e. $\b^T\1 = 0$. Therefore, the only basis vectors with non-zero coefficients
are the ones whose support contains a pair of nodes with different activation values. The number of
node pairs with different activation values can be bounded by the total number of edge flips
(variables that do not agree with their parent variables) in the tree. Let $D_\ell$ denote the
number of edge flips at level $\ell$. Since there are no more than $dL$ basis vectors supported on
a node pair with different activation values, the total number of non-zero coefficients
$\|\B^T\x\|_0 \leq dL \sum_\ell D_\ell$.

Now observe that the tree-structured Ising model essentially specifies that edge flips are
independent and occur with probability $q_\ell = 1/(1+e^{\gamma_\ell}) = 1/(1+d^{\beta\ell})$ at level $\ell$.
That is, the number of flips per level $D_\ell \sim$ Binomial$(|E_\ell|,q_\ell)$ where
$E_\ell \ (= d^\ell)$ denotes the number of edges at level $\ell$. Let $\ell' = L(1-\beta) = (1-\beta)\log_d p$.
Now $d^{\ell(1-\beta)}/2 \leq |E_\ell|q_\ell \leq d^{\ell(1-\beta)}$, and therefore $|E_\ell|q_\ell \rightarrow\infty$ as
$p\rightarrow\infty$ for all $\ell > \ell'$.
Invoking the relative Chernoff bound, we have: For any $\ell > \ell'$, with probability $>1-\delta/L$,
$2^{-1}|E_\ell| q_\ell \leq D_\ell
\leq 2|E_\ell| q_\ell
$
for $p$ large enough.
We can now derive the following bound which holds with probability $>1-\delta$
\begin{eqnarray*}
\|\B^T\x\|_0 & \leq & dL \left(\sum^{\ell'}_{\ell=1}D_\ell + \sum^{L}_{\ell=\ell'+1}D_\ell\right)\\
& \leq & dL \left(\sum^{\ell'}_{\ell=1}|E_\ell| + \sum^{L}_{\ell=\ell'+1}2|E_\ell|q_\ell\right)\\
& \leq & dL \left(\sum^{\ell'}_{\ell=1}d^\ell + \sum^{L}_{\ell=\ell'+1}2d^{\ell(1-\beta)}\right)\\
& \leq & 3dL^2 d^{L(1-\beta)}.
\end{eqnarray*}

\subsection*{Proof of Theorem~\ref{thm:trans_can_sparsity}}
For $\ell < \ell_0$, $\gamma_\ell = \infty$ implies that the probability of edge flip at
level $\ell$, $q_\ell = 0$.
Following the the proof of Theorem~\ref{thm:trans_sparsity},
the bound on the transform domain sparsity still holds.

To evaluate the canonical domain sparsity, we condition on patterns for which
the root variable is zero (inactive). Let $A_\ell$ denote the number of variables
that are active (take value 1) at level $\ell$. Since $q_\ell = 0$ for $\ell < \ell_0$,
there are no flips and hence no variables are active up to level $\ell_0$, i.e. $A_\ell = 0$
for $\ell < \ell_0$. We essentially argue that the canonical sparsity is governed by the
number of nodes that are activated by flips at level $\ell_0$. Flips at lower levels
might activate/de-activate some of the nodes but their effect is insignificant.

First, observe that the number of active variables at level $\ell_0$, conditioned on the root
variable being inactive, is simply the number of edge flips $D_{\ell_0}$ at level $\ell_0$,
i.e. $A_{\ell_0} = D_{\ell_0}$.
Consider $\ell > \ell_0$. Let $M_\ell$ denote the number of active variables at level
$\ell$ whose parents were inactive, and let $N_\ell$ denote the number of active variables
at level $\ell$ whose parents were also active. Therefore, $A_\ell = M_\ell + N_\ell$.
Observe that, conditioned on the values of the variables at level $\ell-1$,
$$M_\ell|A_{\ell-1} \sim \mbox{ Binomial}((|E_{\ell-1}| - A_{\ell-1})d,q_\ell)$$
$$N_\ell|A_{\ell-1} \sim \mbox{ Binomial}(A_{\ell-1}d,1-q_\ell)$$
To gain some understanding for the canonical sparsity, we first look at the
expected canonical sparsity. 
Note that
$\E[\|\x\|_0] = \E[A_L] = \E[\E[A_L|A_{L-1}]] = \E[\E[M_L+N_L|A_{L-1}]]$. 

For the lower bound, we proceed as follows.
$$\E[A_L] \geq \E[\E[N_L|A_{L-1}]] \geq \E[A_{L-1}]d(1-q_L)$$
Now, repeatedly applying similar arguments for $\ell > \ell_0$, we get:
\begin{eqnarray*}
\E[\|\x\|_0] & \geq &\E[A_{\ell_0}] d^{L-\ell_0} \Pi^L_{\ell>\ell_0} (1-q_\ell)\\
& \geq& |E_{\ell_0}|q_{\ell_0} d^{L-\ell_0} (1-q_{\ell_0})^{L-\ell_0}\\
& \geq & \frac{d^{\ell_0(1-\beta)}}{2} d^{L-\ell_0} (1-d^{-\ell_0\beta})^{L-\ell_0}\\
& = & \frac1{2}d^L d^{-\ell_0\beta}(1-p^{-\alpha})^{\log_d p^{1-\frac{\alpha}{\beta}}} \\
& \geq &  cd^L d^{-\ell_0\beta} = c p^{1-\alpha},
\end{eqnarray*}
where $c < 1$. The second step uses the fact that $1-q_\ell$ decreases with $\ell$, and that $A_{\ell_0} = D_{\ell_0} \sim$ Binomial$(|E_{\ell_0}|,q_{\ell_0})$. The last inequality holds for large enough $p$.

For the upper bound, we proceed as follows.
\begin{eqnarray*}
\E[A_L] & = & \E[\E[M_L+N_L|A_{L-1}]] \\
& = & \E[(|E_{L-1}|-A_{L-1})d q_L + A_{L-1}d(1-q_L)] \\
& \leq & |E_{L-1}|d q_L + \E[A_{L-1}]d
\end{eqnarray*}
Repeatedly applying similar arguments for $\ell > \ell_0$, we get:
\begin{eqnarray*}
\E[\|\x\|_0] & \leq & \sum^{L-\ell_0}_{\ell = 1} |E_{L-\ell}|d^\ell q_{L-\ell+1} + \E[A_{\ell_0}]d^{L-\ell_0}\\
& \leq & \sum^{L-\ell_0}_{\ell = 1} d^{L} d^{-(L-\ell+1)\beta} + |E_{\ell_0}|q_{\ell_0} d^{L-\ell_0}\\
& \leq & L d^L d^{-(\ell_0+1)\beta} + d^{\ell_0(1-\beta)}d^{L-\ell_0}\\
& \leq & (L+1) d^L d^{-\ell_0\beta} \leq C (\log_d p) p^{1-\alpha},
\end{eqnarray*}
where $C> 1$. The second step uses the fact that $A_{\ell_0} = D_{\ell_0} \sim$ Binomial$(|E_{\ell_0}|,q_{\ell_0})$.

We now show that similar bounds on canonical sparsity hold with high probability as well.
For this, we will invoke the relative Chernoff bound for binomial random variables $M_\ell$ and $N_\ell$.
First, we derive a lower bound on $A_\ell$ for $\ell > \ell_0$ recursively as follows.
Recall that $A_{\ell_0} = D_{\ell_0} \sim$ Binomial$(|E_{\ell_0}|,q_{\ell_0})$ and
using relative Chernoff bound as in the previous proof, w.p. $>1-\delta/L$, $A_{\ell_0} = D_{\ell_0} \geq \E[D_{\ell_0}]/2
= |E_{\ell_0}|q_{\ell_0}/2 \geq d^{\ell_0(1-\beta)}/4 \rightarrow \infty$ since $\ell_0 = \frac{\alpha}{\beta}L = \frac{\alpha}{\beta}\log_d p\rightarrow\infty$.
Now $A_{\ell_0+1} \geq N_{\ell_0+1}$. And $\E[N_{\ell_0+1}|A_{\ell_0}] = A_{\ell_0}d(1-q_{\ell_0+1}) \geq A_{\ell_0}d(1-q_{\ell_0}) \geq A_{\ell_0}d(1-d^{-\ell_0\beta}) = A_{\ell_0}d(1-p^{-\alpha})$.
Thus, $\E[N_{\ell_0+1}|A_{\ell_0}]\rightarrow\infty$ w.p. $>1-\delta/L$. Conditioning on the values of the
variables at level $\ell_0$ and using relative Chernoff bound, we have with probability $>1-2\delta/L$,
\begin{eqnarray*}
A_{\ell_0+1} \geq N_{\ell_0+1} \geq \E[N_{\ell_0+1}|A_{\ell_0}](1-\epsilon_{\ell_0+1}) \geq A_{\ell_0}d(1-p^{-\alpha})(1-\epsilon_{\ell_0+1})
\end{eqnarray*}
where
\begin{eqnarray*}
\epsilon_{\ell_0+1} & = & \sqrt{\frac{3\log (L/\delta)}{\E[N_{\ell_0+1}|A_{\ell_0}]}}
\leq \sqrt{\frac{3\log (L/\delta)}{A_{\ell_0}d(1-p^{-\alpha})}} \\
& \leq & c' \ p^{-\frac{\alpha}{2\beta}(1-\beta)}\sqrt{\log\log p} <1
\end{eqnarray*}
for $p$ large enough and $c' > 0$ is a constant. 
Notice that $A_{\ell_0+1}\rightarrow \infty$ with probability $>1-2\delta/L$.
Now consider any $\ell > \ell_0$ and assume that for all $\ell \geq \ell' >\ell_0$, $A_{\ell'} \geq A_{\ell'-1}d(1-p^{-\alpha})(1-\epsilon_{\ell'})$,
where $\epsilon_{\ell'} \leq 
c' p^{-\frac{\alpha}{2\beta}(1-\beta)}\sqrt{\log\log p} <1$,
and $A_{\ell'}\rightarrow\infty$ with probability $>1-(\ell'-\ell_0+1)\delta/L$.
We show that similar arguments are true for $A_{\ell+1}$.
Recall that $A_{\ell+1} \geq N_{\ell+1}$. And $\E[N_{\ell+1}|A_{\ell}] = A_{\ell}d(1-q_{\ell+1}) \geq
A_\ell d(1-q_{\ell_0}) \geq A_\ell d(1-p^{-\alpha})$.
Thus, $\E[N_{\ell+1}|A_{\ell}]\rightarrow\infty$ w.h.p. since $A_\ell\rightarrow\infty$. Now, conditioning on the values of the
variables at level $\ell$ and using relative Chernoff bound, we have with probability $>1-(\ell-\ell_0+2)\delta/L$,
\begin{eqnarray*}
A_{\ell+1} & \geq & N_{\ell+1} \geq \E[N_{\ell+1}|A_{\ell}](1-\epsilon_{\ell+1})
\geq A_\ell d(1-p^{-\alpha})(1-\epsilon_{\ell+1})
\end{eqnarray*}
where
\begin{eqnarray*}
\epsilon_{\ell+1} & = & \hspace{-0.3cm}\sqrt{\frac{3\log (L/\delta)}{\E[N_{\ell+1}|A_{\ell}]}}
\leq \sqrt{\frac{3\log (L/\delta)}{A_{\ell}d(1-p^{-\alpha})}} \\
& \leq & \hspace{-0.3cm}\sqrt\frac{3\log (L/\delta)}{A_{\ell_0}(1-p^{-\alpha})^{\ell+1-\ell_0} d^{\ell+1-\ell_0}\Pi^\ell_{\ell'=\ell_0+1}(1-\epsilon_{\ell'}) }\\
& \leq & \hspace{-0.3cm}
c' p^{-\frac{\alpha}{2\beta}(1-\beta)}\sqrt{\log\log p}
\end{eqnarray*}
The last step follows by recalling that $A_{\ell_0}\geq d^{\ell_0(1-\beta)}/4 = p^{\frac{\alpha}{\beta}(1-\beta)}/4$ and
$(1-p^{-\alpha})^{\ell+1-\ell_0}\geq (1-p^{-\alpha})^{L+1-\ell_0} = (1-p^{-\alpha})^{(1-\alpha/\beta)\log_d p+1}>c'$ for large enough $p$.
Also, $\epsilon_{\ell'} \leq 1/2$ for large enough
$p$ and hence $d^{\ell+1-\ell_0}\Pi^\ell_{\ell'=\ell_0+1}
(1-\epsilon_{\ell'}) \geq d(d/2)^{\ell-\ell_0} \geq 1$.

Thus we get, with probability $>1-\delta$, for all $\ell > \ell_0$
\begin{eqnarray*}
A_\ell \geq A_{\ell_0}d^{\ell-\ell_0}(1-p^{-\alpha})^{\ell-\ell_0}\Pi^{\ell}_{\ell'=\ell_0+1}(1-\epsilon_{\ell'})
\end{eqnarray*}
where $\epsilon_{\ell'} \leq c' p^{-\frac{\alpha}{2\beta}(1-\beta)}\sqrt{\log\log p}<1$.
Finally, we have a lower bound on the canonical sparsity as follows:
With probability $>1-\delta$,
\begin{eqnarray*}
\|\x\|_0 = A_L & \geq & A_{\ell_0}d^{L-\ell_0}((1-p^{-\alpha})(1-c' \ p^{-\frac{\alpha(1-\beta)}{2\beta}}\log p))^{L-\ell_0} \\
& \geq & c d^{\ell_0(1-\beta)} d^{L-\ell_0} = c d^L d^{-\ell_0\beta} = c p^{1-\alpha}
\end{eqnarray*}
where we use the fact that $(1-p^{-a})^{\log_d p^b} \geq c>0$ for large enough $p$. Also note that $c<1$.

We now establish an upper bound on the canonical sparsity. Recall that $A_\ell = M_\ell + N_\ell$.
In the analysis above, we
established that $\E[N_\ell|A_{\ell-1}] \rightarrow\infty$ for each $\ell > \ell_0$ w.p. $>1-\delta/L$.
Now consider $M_\ell$. We show that $\E[M_\ell|A_{\ell-1}] 
\rightarrow\infty$ w.p. $> 1-\delta/L$, 
and derive an upper bound on $A_\ell$ for $\ell >\ell_0$ recursively as follows.
Recall that $A_{\ell_0} = D_{\ell_0} \sim$ Binomial$(|E_{\ell_0}|,q_{\ell_0})$ and
using relative Chernoff bound as in the previous proof, w.p. $>1-\delta/L$, $A_{\ell_0} = D_{\ell_0} \leq 2\E[D_{\ell_0}]
= 2|E_{\ell_0}|q_{\ell_0}$. Now $\E[M_{\ell_0+1}|A_{\ell_0}]
= (|E_{\ell_0}|-A_{\ell_0})d q_{\ell_0+1} \geq |E_{\ell_0}|(1-2q_{\ell_0})d q_{\ell_0+1} \geq d^{(\ell_0+1)(1-\beta)}(1-2d^{-\ell_0\beta})/2 = d^{(\ell_0+1)(1-\beta)}(1-2p^{-\alpha})/2\rightarrow \infty$ since $\ell_0 = \frac{\alpha}{\beta}L = \frac{\alpha}{\beta}\log_d p \rightarrow \infty$. Thus, $\E[M_{\ell_0+1}|A_{\ell_0}]\rightarrow\infty$ w.p. $>1-\delta/L$.
Conditioning on the values of the variables at level $\ell_0$ and using relative Chernoff bound,
we have with probability $>1-4\delta/L$,
\begin{eqnarray*}
A_{\ell_0+1} & = & N_{\ell_0+1} + M_{\ell_0+1} \\
& \leq & (1+\epsilon_{\ell_0+1})(\E[N_{\ell_0+1}|A_{\ell_0}] + \E[M_{\ell_0+1}|A_{\ell_0}])\\
& \leq & (1+\epsilon_{\ell_0+1})(A_{\ell_0} + |E_{\ell_0}|q_{\ell_0+1})d
\end{eqnarray*}
where
\begin{eqnarray*}
\epsilon_{\ell_0+1} & = & \max\left(\sqrt{\frac{3\log (L/\delta)}{\E[N_{\ell_0+1}|A_{\ell_0}]}},\sqrt{\frac{3\log (L/\delta)}{\E[M_{\ell_0+1}|A_{\ell_0}]}}\right)\\
& \leq &\max\left(\sqrt{\frac{3\log (L/\delta)}{A_{\ell_0}d(1-p^{-\alpha})}},\sqrt{\frac{6\log (L/\delta)}{d^{(\ell_0+1)(1-\beta)}(1-2p^{-\alpha})}}\right) \\
& \leq & c' \ p^{-\frac{\alpha}{2\beta}(1-\beta)}\sqrt{\log\log p}<1
\end{eqnarray*}
for $p$ large enough and $c' > 0$ is a constant.
Now consider any $\ell > \ell_0$ and assume that for all $\ell \geq \ell' >\ell_0$,
with probability $>1-2(\ell'-\ell_0+1)\delta/L$, $\E[M_{\ell'}|A_{\ell'-1}]\rightarrow\infty$
and
$$A_{\ell'} \leq (1+\epsilon_{\ell'})(A_{\ell'-1} + |E_{\ell'-1}|q_{\ell'})d,$$
where $\epsilon_{\ell'} \leq 
c' p^{-\frac{\alpha}{2\beta}(1-\beta)}\sqrt{\log\log p}<1$.
We show that similar arguments are true for $A_{\ell+1}$.
Recall that $A_{\ell+1} = N_{\ell+1} + M_{\ell+1}$. Using the upper bound on $A_{\ell'}$ for $\ell \geq \ell' > \ell_0$
recursively, we have with probability $>1-2(\ell-\ell_0+1)\delta/L$,
\begin{eqnarray*}
\E[M_{\ell+1}|A_{\ell}] & = & (|E_\ell|-A_{\ell})d q_{\ell+1}\\
& \geq & |E_\ell| d q_{\ell+1} - (1+\epsilon_{\ell})(A_{\ell-1} + |E_{\ell-1}|q_{\ell})d^2 q_{\ell+1}\\
& \geq & |E_\ell| d q_{\ell+1} - \sum^\ell_{\ell'=\ell_0+1} |E_{\ell'-1}|q_{\ell'}d^{\ell+2-\ell'} q_{\ell+1} \Pi^\ell_{\ell'' = \ell'}
(1+\epsilon_{\ell''}) \\
& &  - A_{\ell_0}d^{\ell-\ell_0+1}q_{\ell+1}\Pi^\ell_{\ell'=\ell_0+1}(1+\epsilon_{\ell'})\\
& \geq & d^{(\ell+1)(1-\beta)}\left[\frac1{2} - \sum^\ell_{\ell'=\ell_0+1}
d^{-\ell'\beta}\Pi^\ell_{\ell'' = \ell'} (1+\epsilon_{\ell''}) - 2d^{-\ell_0\beta}\Pi^\ell_{\ell'=\ell_0+1}(1+\epsilon_{\ell'})\right]\\
& \geq & d^{(\ell+1)(1-\beta)}\left[\frac1{2} - 3Ld^{-\ell_0\beta} (1+c' p^{-\frac{\alpha}{2\beta}(1-\beta)}\log p)^{\ell-\ell_0} \right]\\
& \geq &  d^{(\ell+1)(1-\beta)}\left[\frac1{2} -3Lcp^{-\alpha}\right] \\
& \geq &  c_\delta d^{(\ell+1)(1-\beta)}\rightarrow \infty
\end{eqnarray*}
The second last line uses the fact that $\ell - \ell_0 \leq L-\ell_0 = (1-\frac{\alpha}{\beta})\log_d p$ and $(1+p^{-a})^{\log_d p^b} \leq e^{p^{-a}\log_d p^b} \leq c$, a constant, for $p$ large enough.  The last step follows for large enough $p$ and since $\ell >\ell_0 \frac{\alpha}{\beta}L = \frac{\alpha}{\beta}\log_d p\rightarrow \infty$.
Thus, $\E[M_{\ell+1}|A_{\ell}]\rightarrow\infty$ w.h.p. Now, conditioning on the values of the
variables at level $\ell$ and using relative Chernoff bound, we have with probability $>1-2(\ell-\ell_0+2)\delta/L$,
\begin{eqnarray*}
A_{\ell+1} & = & N_{\ell+1}+M_{\ell+1} \\
& \leq & (1+\epsilon_{\ell+1})(\E[N_{\ell+1}|A_{\ell}]+\E[M_{\ell+1}|A_{\ell}]) \\
& \leq & (1+\epsilon_{\ell+1})(A_{\ell} + |E_{\ell}|q_{\ell+1})d
\end{eqnarray*}
where
\begin{eqnarray*}
\epsilon_{\ell+1} & = & \max\left(\sqrt{\frac{3\log (L/\delta)}{\E[N_{\ell+1}|A_{\ell}]}},\sqrt{\frac{3\log (L/\delta)}{\E[M_{\ell+1}|A_{\ell}]}}\right)\\
& \leq &\max\left(\sqrt{\frac{3\log (L/\delta)}{A_{\ell}d(1-p^{-\alpha})}},\sqrt{\frac{6\log (L/\delta)}{d^{(\ell+1)(1-\beta)}(1-6Lcp^{-\alpha})}}\right) \\
& \leq & c' \ p^{-\frac{\alpha}{2\beta}(1-\beta)}\sqrt{\log\log p}<1
\end{eqnarray*}
for $p$ large enough.

Thus using recursion we get, with probability $>1-2\delta$, for all $\ell > \ell_0$
\begin{eqnarray*}
A_\ell & \leq & A_{\ell_0}d^{\ell-\ell_0}\Pi^\ell_{\ell' = \ell_0+1} (1+\epsilon_\ell')
+ \sum^\ell_{\ell'= \ell_0+1}|E_{\ell'-1}|q_{\ell'} d^{\ell-\ell'+1} \Pi^\ell_{\ell'' = \ell'}(1+\epsilon_{\ell''})\\
& \leq & 2d^{-\ell_0\beta} d^\ell \Pi^\ell_{\ell' = \ell_0+1} (1+\epsilon_\ell') + \sum^\ell_{\ell'= \ell_0+1}d^{\ell}d^{-\ell'\beta} \Pi^\ell_{\ell'' = \ell'}(1+\epsilon_{\ell''})\\
& \leq & Cd^\ell d^{-\ell_0\beta}
\end{eqnarray*}
where $C>1$ is a constant.
Last step uses the fact that $\epsilon_{\ell} \leq c' p^{-\frac{\alpha}{2\beta}(1-\beta)}\sqrt{\log\log p}$, and $(1+p^{-a})^{\log_d p^b} \leq e^{p^{-a}\log_d p^b} \leq c$, a constant, for $p$ large enough.
Finally, we have an upper bound on the canonical sparsity as follows:
With probability $>1-2\delta$,
$$
\|\x\|_0 =A_L \leq Cd^L d^{-\ell_0\beta} = C p^{1-\alpha}.
$$

\subsection*{Proof of Theorem~\ref{thm:detect}}
Consider the threshold $t = \sqrt{2\sigma^2(1+c)\log p}$, where $c>0$ is an arbitrary constant.
Since the proposed transform is orthonormal, it is easy to see that under the null hypothesis
$H_0$ (no activation), the empirical transform coefficients $\b^T_i \y \sim \N(0,\sigma^2)$. Therefore,
the false alarm probability can be bounded as follows:
\begin{eqnarray*}
P_{H_0}(\max_i |\b_i^T\y| > t) 
= 1-\Pi^p_{i=1}P_{H_0}(|\b_i^T\y| \leq t)
& \leq & 1 - (1-2e^{-t^2/2\sigma^2})^p\\
& = & 1-\left(1-\frac1{p^{1+c}}\right)^p\rightarrow 0
\end{eqnarray*}
Under the alternate hypothesis $H_1$ ($\x \neq 0$), the empirical transform coefficients $\b^T_i \y \sim \N(\mu\b^T_i \x,\sigma^2)$.
Therefore, the miss probability can be bounded as follows:
\begin{eqnarray*}
P_{H_1}(\max_i |\b_i^T\y| \leq t) 
& \leq &\Pi_{i:\b_i^T\x \neq 0} P(|\N(\mu\b_i^T\x,\sigma^2)| \leq t)\\
&\leq & \Pi_{i:\b_i^T\x > 0} P(\N(\mu\b_i^T\x,\sigma^2) \leq t) \cdot \Pi_{i:\b_i^T\x < 0} P(\N(\mu\b_i^T\x,\sigma^2) \geq -t)\\
& = & \Pi_{i:\b_i^T\x > 0} P(\N(0,\sigma^2) \leq t-\mu|\b_i^T\x|) \cdot \Pi_{i:\b_i^T\x < 0} P(\N(0,\sigma^2) \geq -t+\mu|\b_i^T\x|)\\
& = & \Pi_{i:\b_i^T\x \neq 0} P(\N(0,\sigma^2) \leq t-\mu|\b_i^T\x|)\\
& \leq & P(\N(0,\sigma^2) \leq t-\mu\max_i|\b_i^T\x|)
\end{eqnarray*}
In the second step we use the fact that $P(|a|\leq t) \leq P(a\leq t)$ and also $P(|a|\leq t) \leq P(a\geq -t)$.
Thus, the miss probability goes to zero if $\mu\max_i|\b_i^T\x| > (1+c')t$ for any arbitrary $c'>0$.

The detectability threshold now follows by deriving a lower bound for the largest absolute transform
coefficient. We employ the simple fact that the energy in the largest transform coefficient is
at least as large as the average energy per non-zero coefficient:
$$
\max_i |\b^T_i \x| \geq \sqrt{\|\x\|_0/\|\B^T\x\|_0}
$$
Now invoking Theorem~\ref{thm:trans_can_sparsity} for patterns
that correspond to the root value zero,
with probability $>1-2\delta$,
$$
\max_i |\b^T_i \x| \geq c \ p^{(\beta-\alpha)/2}
$$
where $c > 0$ is a constant.  Patterns that do not correspond to
the root variable taking value zero are canonically non-sparse and have
$\|\x\|_0$ larger than the patterns that correspond to the root variable taking value zero.
Therefore, the same lower bound holds in this case as well.

\subsection*{Proof of Theorem~\ref{thm:emp_cov}}
Observe that the true hierarchical structure $\H^*$ between the leaf variables
can be recovered if the empirical covariances $\{\widehat r_{ij}\}$ satisfy the
conditions of Lemma~\ref{lem:clus}. Recall that $\{\E[(x_i x_j)]\}$ satisfy the
conditions of Lemma~\ref{lem:clus}, and the true covariance of the observed
variables $r_{ij} = \E[(y_i y_j)]=\E[(x_i x_j)]$ for $i\neq j$ (the auto-covariances
are not important for clustering).  Also, recall that $\tau$ denotes the smallest difference
(gap) between the minimum pairwise covariance of leaf variables within any cluster
and the maximum covariance between leaf variables in different clusters.
Hence, a sufficient condition for the empirical
covariances $\{\widehat r_{ij}\}$ to satisfy the conditions of Lemma~\ref{lem:clus}
is that the deviation between true and empirical covariance of the observed variables
is less than $\tau/2$, i.e.
\begin{equation}
\label{eq:cov_dev}
\max_{(i,j)}|\widehat r_{ij}-r_{ij}|< \tau/2.
\end{equation}

To establish Eq.~\ref{eq:cov_dev}, we study the concentration of the empirical covariances
around the true covariances. For this, we first argue that the random variable
$v_k := y^{(k)}_iy^{(k)}_j$ satisfies the following moment conditions:
$$
\E[|v_k-\E[v_k]|^p] \leq \frac{p! \mbox{var}(v_k)h^{p-2}}{2}
$$
for integers $p\geq 2$ and some constant $h>0$. We will make use the following three
results (Lemmas 1-3 from \cite{Randproj_JHaupt}):
\begin{itemize}
\item[1)] If the even absolute central moments of a random variable satisfy the moment
condition, then so do the odd moments. This implies that Gaussian random variables satisfy
moment conditions since the even moments of $A\sim\N(\mu,\sigma^2)$ are given as
$$
\E[|A-\mu|^{2p}] = 1.3.5.\dots.(2p-1) \sigma^{2p}.
$$
\item[2)] If two zero-mean random variables $(A,B)$ satisfy the moment conditions and
$\E[AB]\geq 0$, then $A+B$ also satisfies the moment condition.
\item[3)] If two zero-mean, independent random variables $(A,B)$ satisfy the moment
conditions, then $AB$ also satisfies the moment condition.
\end{itemize}
Now observe that
\begin{eqnarray*}
v_k & = & (x^{(k)}_i+\epsilon^{(k)}_i)(x^{(k)}_j+\epsilon^{(k)}_j)\\
& = & x^{(k)}_i x^{(k)}_j + x^{(k)}_i\epsilon^{(k)}_j + \epsilon^{(k)}_i x^{(k)}_j
+ \epsilon^{(k)}_i\epsilon^{(k)}_j.
\end{eqnarray*}
We will now argue that each of the terms in the above expression satisfy moment conditions.
Since $|x^{(k)}|,|x^{(k)}_j|$ are bounded, $x^{(k)}_i, x^{(k)}_j$ as well as the first term
$x^{(k)}_i x^{(k)}_j$ satisfy the moment condition. Also, since $\epsilon^{(k)}_i,\epsilon^{(k)}_j$
are gaussian, they satisfy the moment conditions as per result 1). And using result 3) above
for the product of independent random variables, we see that the remaining three terms
$x^{(k)}_i \epsilon^{(k)}_j, \epsilon^{(k)}_i x^{(k)}_j, \epsilon^{(k)}_i\epsilon^{(k)}_j$
satisfy the moment conditions. Now it is not too hard to see that for any two terms $A,B$ in
the expression above, $\E[AB]\geq 0$. Therefore, using result 2) above for the sum of random
variables, we get that $v_k$ satisfies the moment condition with some parameter $h$.
Also, since $\{v_k\}^n_{k=1}$ are independent, we can now invoke the Bernstein inequality to get:
\begin{eqnarray*}
P\left(\frac1{n}\sum^n_{k=1}(v_k - \E[v_k])> \frac{2t}{n}\sqrt{\sum^n_{k=1}\mbox{var}(v_k)}\right) < e^{-t^2}
\end{eqnarray*}
for $0<t\leq \sqrt{\sum^n_{k=1}\mbox{var}(v_k)}/(2h)$.
Now, straight-forward computations show that
$$
\mbox{var}(v_k) = \left\{
\begin{array}{c l}
\sigma^4+\sigma^2\left(\E\left[(x^{(k)}_i)^2\right]+\E\left[(x^{(k)}_j)^2\right]\right) + \mbox{var}\left(x^{(k)}_i x^{(k)}_j\right) & i\neq j\\
2\sigma^4+4\sigma^2\E\left[(x^{(k)}_i)^2\right]+\mbox{var}\left((x^{(k)}_i)^2\right) & i=j
\end{array}
\right.
$$
Since $|x^{(k)}_i|\leq M$, we have $c_1:= \sigma^4 \leq \mbox{var}(v_k) \leq 2\sigma^4+4M^2\sigma^2+4M^4 =:c_2$. And we get
\begin{eqnarray*}
P\left(\frac1{n}\sum^n_{k=1}(v_k - \E[v_k])> \frac{2t\sqrt{c_2}}{\sqrt{n}}\right) < e^{-t^2}
\end{eqnarray*}
Let $t = \sqrt{n}\tau/(4\sqrt{c_2\log n})$,
where $\tau$ is the gap between the minimum pairwise covariance of variables within any cluster
and the maximum covariance between variables in different clusters. Then we get:
\begin{eqnarray*}
P\left(\frac1{n}\sum^n_{k=1}(v_k - \E[v_k])> \frac{\tau}{2}\right) < e^{-n\tau^2/(16c_2\log n)}
\end{eqnarray*}
and $0<t = \sqrt{n}\tau/(4\sqrt{c_2\log n}) \leq \sqrt{n c_1}/(2h) \leq \sqrt{\sum^n_{k=1}\mbox{var}(v_k)}/(2h)$
for large enough $n$ and hence $t$ satisfies the desired conditions. Similar arguments show that
$-v_k$ also satisfies the moment condition, and hence we get:
\begin{eqnarray*}
P\left(\left|\frac1{n}\sum^n_{k=1}(v_k - \E[v_k])\right|\geq \frac{\tau}{2}\right) < 2e^{-n\tau^2/(16c_2\log n)}
\end{eqnarray*}
Equivalently,
$$
P(|\widehat r_{ij}-r_{ij}|> \tau/2)< 2e^{-n\tau^2/(16c_2\log n)}
$$
And taking union bound over all elements in the similarity matrix, we have that the
\begin{eqnarray*}
P(\max_{ij}|\widehat r_{ij}-r_{ij}|> \tau/2) < 2 p^2 e^{-n \tau^2/(16c_2\log n)}.
\end{eqnarray*}
Thus, the covariance clustering algorithm of Figure~\ref{tab:clustering_algo}
recovers $\H^*$ with probability $>1-\delta$ from
$$
\frac{n}{\log n} \geq \frac{16c_2}{\tau^2}\log(2 p^2/\delta)
$$
i.i.d snapshots of leaf variables.

%
%
%
%
%
%
%
%
%
%
%
%
%

\bibliographystyle{IEEEtran}
\small \bibliography{thesis}

\end{document}